\documentclass[aps,pre,twocolumn,superscriptaddress,showpacs]{revtex4}
\usepackage{amssymb}
\usepackage{epsfig}
\usepackage{amsmath}
\usepackage{times}

\begin{document}

\title{Onset of Synchronization in Weighted Complex Networks: \\
the Effect of Weight-Degree Correlation}

\author{Menghui Li}
\affiliation{Temasek Laboratories, National University of
Singapore, 117508, Singapore} \affiliation{Beijing-Hong
Kong-Singapore Joint Centre for Nonlinear \& Complex Systems
(Singapore), National University of Singapore, Kent Ridge, 119260,
Singapore}
\author{Xingang Wang}\email[Corresponding author. Email address: ]{wangxg@zju.edu.cn}
 \affiliation{Institute for
Fusion Theory and Simulation, Zhejiang University, Hangzhou 310027,
China }
\author{Ying Fan}
\affiliation{Department of Systems Science, School of Management, Beijing Normal University, Beijing 100875, China}
\author{Zengru Di}
\affiliation{Department of Systems Science, School of Management, Beijing Normal University, Beijing 100875, China}
\author{Choy-Heng Lai}
\affiliation{Beijing-Hong Kong-Singapore Joint Centre for
Nonlinear \& Complex Systems (Singapore), National University of
Singapore, Kent Ridge, 119260, Singapore} \affiliation{Department
of Physics, National University of Singapore, Singapore 117542}

\begin{abstract}
By numerical simulations, we investigate the onset of
synchronization of networked phase oscillators under two different
weighting schemes. In scheme-I, the link weights are correlated to
the product of the degrees of the connected nodes, so this kind of
networks is named as the weight-degree correlated (WDC) network. In
scheme-II, the link weights are randomly assigned to each link
regardless of the node degrees, so this kind of networks is named as
the weight-degree uncorrelated (WDU) network. Interestingly, it is
found that by increasing a parameter that governs the weight
distribution, the onset of synchronization in WDC network is
monotonically enhanced, while in WDU network there is a reverse in
the synchronization performance. We investigate this phenomenon from
the viewpoint of gradient network, and explain the contrary roles of
coupling gradient on network synchronization: gradient promotes
synchronization in WDC network, while deteriorates synchronization
in WDU network. The findings highlight the fact that, besides the
link weight, the correlation between the weight and node degree is
also important to the network dynamics.

\end{abstract}
\date{\today }
\pacs{05.45.Xt, 89.75.Hc} \maketitle

\textbf{In the past decades, complex networks have became an
established framework to understand the behavior of a large variety
of complex systems, ranging from biology to social sciences
\cite{Review-RMP,Review-SIAM,Review-weighted}. In network studies,
one important issue is to investigate the influence of the network
topology on the dynamics, e.g., the synchronization behaviors of
coupled oscillators \cite{SyninNetworks}. For many realistic
systems, besides the degree, the weight of the network links also
presents the heterogeneous distribution, i.e., the weighted
networks. Depending on the correlation between the node degree and
the link weight, the weighted systems can be roughly divided into
two types: weight-degree correlated (WDC) and weight-degree
uncorrelated (WDU) networks \cite{Review-weighted}. While the
collective dynamics taking place on weighted networks have been
extensively studied \cite{SyninNetworks}, but so far, to the best of
our knowledge, the impact of weight-degree correlation on the
collective dynamics has not been addressed in literature. In this
paper, we consider the effect of weight-degree correlation on the
onset of synchronization of a generalized Kuramoto model with fixed
total coupling cost. Interestingly, it is found that, by increasing
the weight parameter that characterizes the distribution of links
weights, the synchronization is monotonically enhanced in WDC
network, while the synchronization could be deteriorated firstly and
then enhanced in WDU network. Moreover, we explain qualitatively the
fundamental mechanism from the viewpoint of gradient network. Our
findings may be helpful to the design and optimization of realistic
dynamical networks.}

\section{Introduction}
Due to its broad applications, synchronization of coupled phase
oscillators has been extensively studied in the past decades
\cite{Kuramoto,JA:2005}. Recently, with the blooming of network
science, this study has been extended to the situation of complex
network structures, i.e., the generalized Kuramoto model
\cite{MP:2004,TI:2004,ROH:Pre,ROH:2005,JYA:2007,GSG:Pre,XGW:Chaos,TA:2009,LGL:2010}.
Different to the traditional studies of regular networks, in the
generalized Kuramoto model more attention has been paid to the
interplay between the network topology and dynamics. Considering the
fact that links in realistic networks are generally of different
strength
\cite{YOOK:2001,NEWMAN:2001a,NEWMAN:2001b,Gomez-Gardenes:2010}, more
recently the study of onset of synchronization in weighted complex
networks has arisen certain interest, where a number of new
phenomena have been identified
\cite{MOTTER:EPL,MOTTER:PRE,HCAB:2005,CHAHB:2005,ZCS:2006,XGW:2007,XGW:2008,SKHJ:2009,WXWang:2009,ZT:2010}.

For weighted networks in reality, a typical feature is that the
weights of the network links possess a large variation. In
particular, it is found that in many realistic networks the link
weights roughly follow power-law distributions, say, for instance,
the airport network \cite{PNAS_101_3747}, the scientific
collaboration network \cite{Physa_375_355}, and the mobile
communication network \cite{PNAS_104_7332}, to name just a few. In
the previous studies of weighted networks, the heterogeneity of link
weights has been attributed to the heterogenous network structure,
e.g., the heterogenous degree distribution, and the weight of a link
in general is set to be proportional to the product of the degrees
of the nodes it connects. For this feature, we call this type of
network as the weight-degree correlated (WDC) network, which is
often observed in the biological and technological systems, e.g.,
the E. Coli metabolic network \cite{EPL_72_308} and the airport
network \cite{EPL_72_308, PNAS_101_3747}. Lately, with the progress
of network science, it is also found that some realistic networks,
while the link weights also possess a power-law distribution, the
link weight and node degree are not closely correlated, e.g., the
scientific collaboration and email networks
\cite{PNAS_101_3747,Physa_378_583}. More specifically, in this type
of networks the link weights seem to be randomly chosen within a
certain range, regardless of the node degrees. For this property, we
call this type of network as the weight-degree uncorrelated (WDU)
network. Our main task in the present paper is to investigate the
dynamical behaviors of these two different types of weighted
networks, with the purpose of exploring the effect of weight-degree
correlation on the network dynamics.

To study the effect of weight-degree correlation on network
dynamics, we will employ the model of networked phase oscillators,
i.e., the generalized Kuramoto model, and study the onset of
synchronization between the following two weighting schemes. In
scheme-I, the coupling weight between a pair of directly connected
nodes in the network is set to be proportional to the degrees of
these two nodes, so as to generate a WDC-type network. In scheme-II,
the weights of the network links also have a power-law distribution,
but the weights are randomly distributed among the links, regardless
of the node degrees. This gives rise to a WDU-type network. Our main
finding is that, by tuning a weight parameter (to be defined later)
that governs the heterogeneity of the link weights, the two types of
networks have very different synchronization responses. In
particular, in WDC-type network, with the increase of the weight
parameter, the network synchronization is found to be
\emph{monotonically} enhanced; while in WDU-type network, it is
found that the increase of the weight parameter may \emph{either
suppress or enhance} the network synchronization. We analyze the
different synchronization responses of the two networks from the
viewpoint of gradient network \cite{GN:2004,XGW:2007}, and get the
general conclusion that synchronization is enhanced by coupling
gradient in WDC network, while is suppressed in WDU network. These
findings highlight that, besides the link weight, the correlation
between the link weight and the node degree should be also taken
into account when analyzing the network dynamics.

The rest of this paper is organized as follows. In Sec. \ref{model},
we will introduce our model of coupled phase oscillators, as well as
the two types of weighting schemes. In Sec. \ref{Phase}, we will
show the transition paths from non-synchronous to synchronous states
for the two types of networks, in which the critical coupling
strength characterizing onset of synchronization will be defined. In
Sec. \ref{criticalcouple}, we will study in detail the effect of
weight-degree correlation on the network synchronization, in which
the reversed synchronization in WDU network will be highlighted.
Also, based on the concept of gradient complex network, we will give
an analysis to the synchronization performance of the two types of
networks. In Sec. \ref{conclud}, we will give our discussion and
conclusion.

\section{The model}\label{model}
In our study, we employ the following two typical models for the
network structure: the small-world and scale-free networks. The
small-world network is constructed by the method introduced in Ref.
\cite{Nature_393_440}, in which each node has $2m$ links and each
link is rewired randomly with a probability $p_r$. The scale-free
network is generated by the standard BA growth model
\cite{Science_286_509}, which has average degree $\langle k\rangle =
2m$, and the node degrees roughly follow  a power-law distribution,
$P(k)\sim k^{-\gamma}$, with $\gamma=3$.

For illustration, we set the weights in the two types of networks as
follows. For the WDC network, motivated by the empirical observation
in metabolic networks \cite{EPL_72_308} and airport networks
\cite{PNAS_101_3747,EPL_72_308}, we set the weight, $c_{mn}$,
between the pair of directly connected nodes, $m$ and $n$, to be
\begin{equation}
c_{mn}\sim (k_mk_n)^{\alpha}, \label{alpha}
\end{equation}
where $k_m$ and $k_n$ are the degrees of the nodes $m$ and $n$,
respectively, and $\alpha\in \mathbb{R}$ is a tunable parameter
characterizing the weight distribution, i.e. the weight parameter.
For the WDU network, motivated by the empirical observation in
scientific collaboration networks\cite{PNAS_101_3747} and email
networks\cite{Physa_378_583}, the weight of each link is randomly
chosen within the range $[c_{min},c_{max}]$, while keeping the
weight distribution to be having the power-law scaling,
\begin{equation}
P(c)\sim c^{-\beta}. \label{powerlaw}
\end{equation}
In WDU network, the weight parameter is $\beta$, which, according to
the empirical observations, is defined to be positive.

Arranging each node in the network with a phase oscillator and
regarding the weight as the coupling strength between the connected
nodes, we now can study the dynamical behaviors of the two types of
networks constructed above. Our model of coupled phase oscillators
reads
\begin{equation}
\dot{\phi}_m = \omega_m + \frac{\varepsilon}{s_m}\sum_{n=1}^N
c_{mn}\sin(\phi_n-\phi_m),\label{Kuramoto}
\end{equation}
with $N$ the network size, and $m,n=1,\ldots,N$ the node indices.
$\phi_m$ and $\omega_m$ represent, respectively, the phase state and
the intrinsic frequency of the $m$th oscillator (node). The weights
of the network links are represented by the matrix $\{c_{mn}\}$,
with $c_{mn}=0$ if $m$ and $n$ are not directly connected.
$\varepsilon$ is a uniform coupling strength which, for the sake of
convenience, is normalized by the node weight intensity
$s_m=\sum_{n=1}^N c_{mn}$. Our main task is to investigate how the
weighting schemes, i.e. weight-degree correlated or non-correlated,
will affect the network dynamics.

In next simulations, we fix the network size as $N=6400$ and the
average degree as $\langle k \rangle =2m=6$. In constructing the
small-world network, we set the rewiring probability $p_r$ as $0.2$.
For each network example, the intrinsic frequency, $\{\omega_i\}$,
of the network oscillators is randomly chosen from the Gaussian
distribution
$g(\omega)=(2\pi\sigma^2)^{-1/2}\exp(-\omega^2/2\sigma^2)$, which
has zero mean value and the unit variance ($\sigma^2=1$). Also, the
initial states of the oscillators are randomly chosen from the
interval $[0,2\pi)$. With these settings, we then set a value for
the uniform coupling strength, $\varepsilon$, and evolve the network
according to Eq. (\ref{Kuramoto}). The equations are integrated by
the fourth-order Runge-Kutta method and the time step is $\Delta
t=0.01$. After discarding a transient process of $2000$ time steps,
we start to measure the collective behavior of the network, which is
characterized by the network order parameter
\cite{ROH:Pre,XGW:Chaos}
\begin{equation}
R= [\langle | \frac{\sum_{n=1}^{N}d_ne^{i\phi_n}}{\sum_{m=1}^{N}
d_m}|\rangle].\label{order}
\end{equation}
Here $\langle\ldots\rangle$ represents that the result is averaged
over a period of $2000$ time steps, and $[\ldots]$ represents the
average of 20 network runs. $d_n$ is the total incoming coupling
strength of the node $n$, which, due to the normalization operation
in the model, is unit. By scanning the coupling strength
$\varepsilon$, we can monitor the behavior of $R$, so as to find the
critical coupling strength, $\varepsilon_c$, where the onset of
network synchronization occurs.

\section{Onset of synchronization in weighted complex networks}\label{Phase}

\begin{figure}
\includegraphics[width=0.85\linewidth]{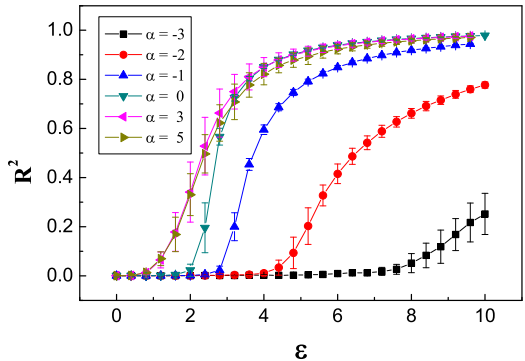}
\caption{(Color online) Under the WDC scheme, the variation of the
synchronization order parameter, $R^2$, as a function of the
coupling strength, $\varepsilon$, for different weight parameters,
$\alpha$. The network is generated by the standard BA model
(scale-free network), which has size $N=6400$ and average degree
$\langle k \rangle =6$. Onset of synchronization is identified as
the coupling strength where $R^2$ starts to increase from $2\times
10^{-2}$. The error bars are estimated by the standard deviation.
Each data is averaged over 20 network runs.} \label{fig1}
\end{figure}

\begin{figure}
\includegraphics[width=0.85\linewidth]{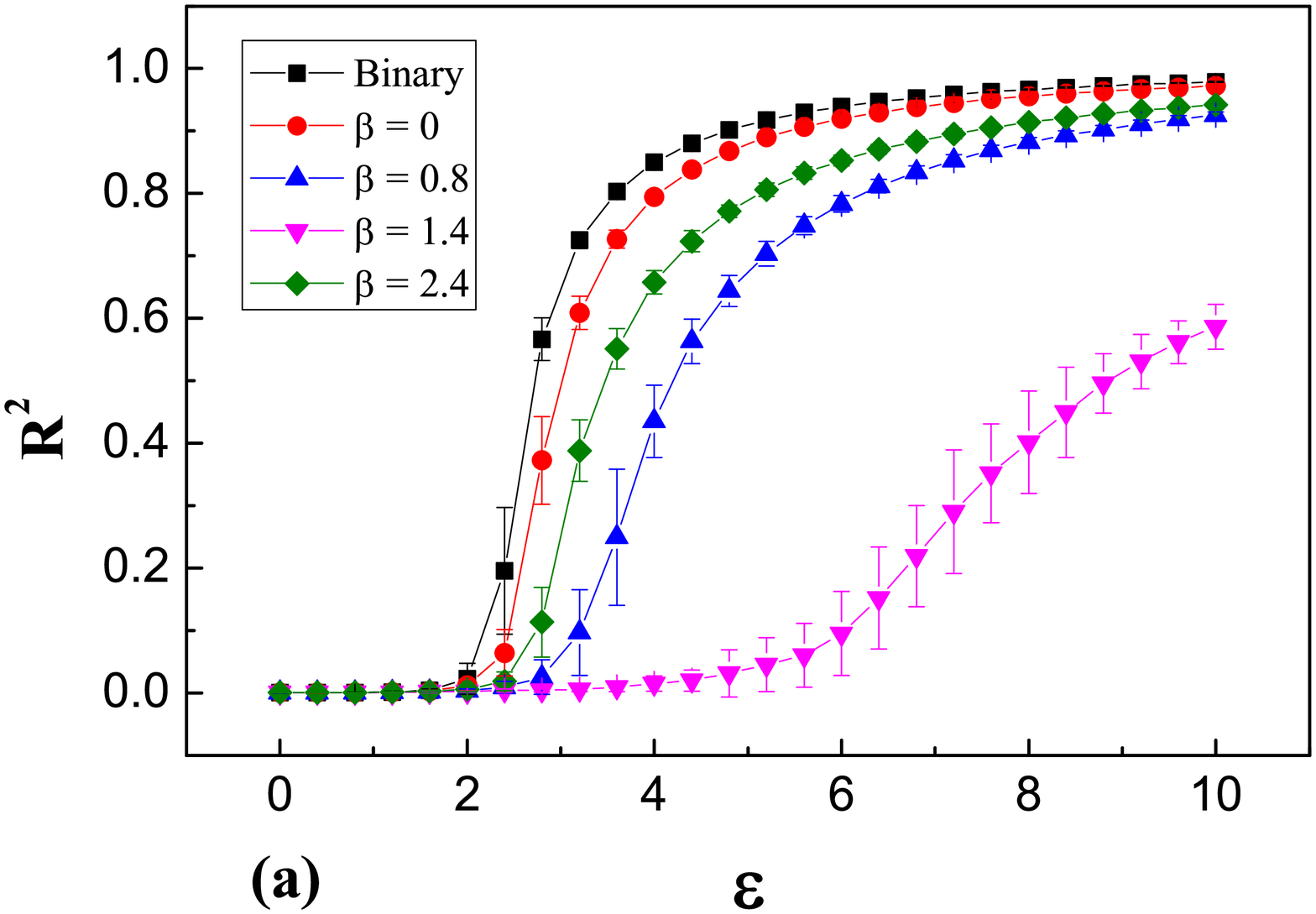}
\includegraphics[width=0.85\linewidth]{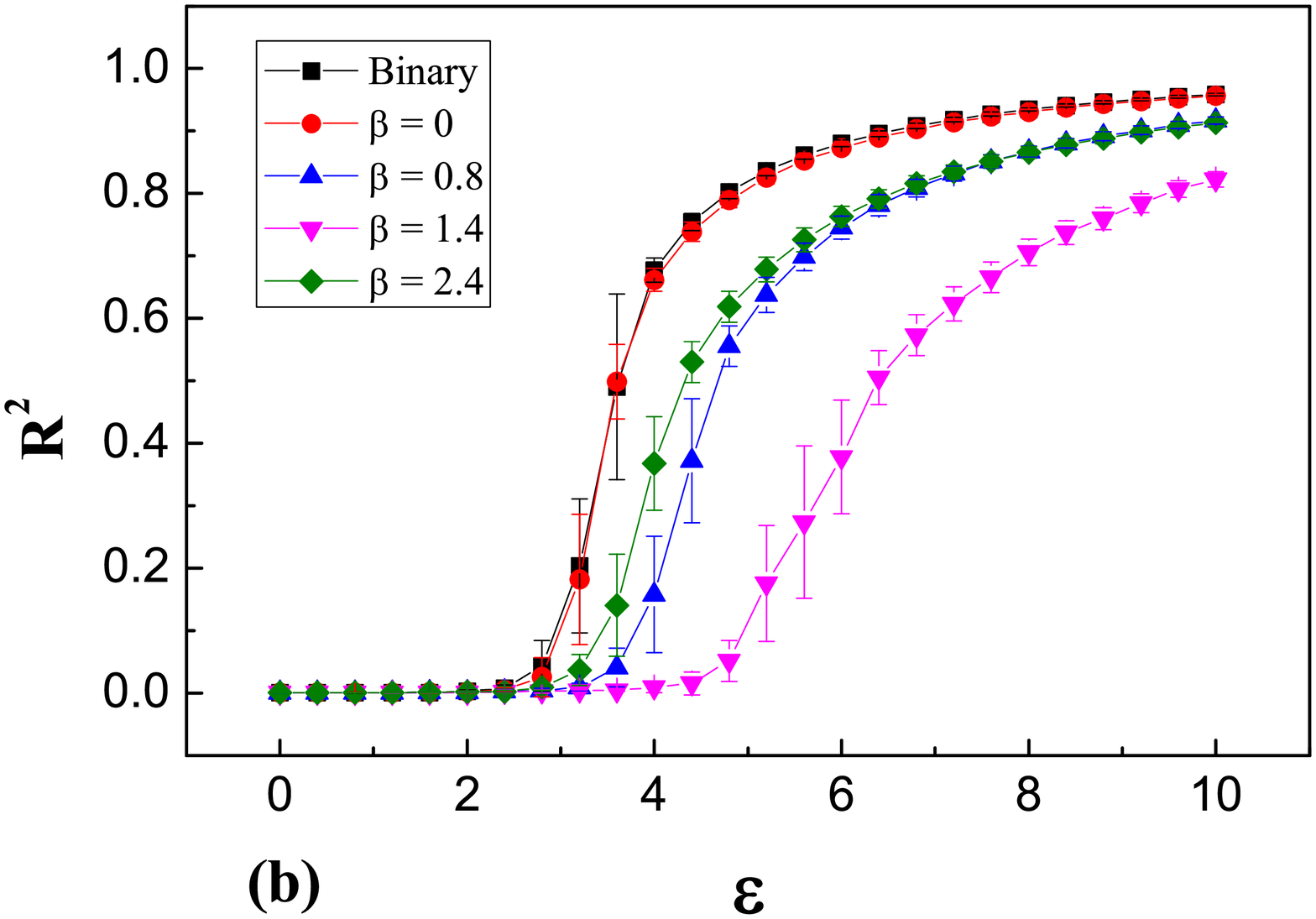}
\includegraphics[width=0.85\linewidth]{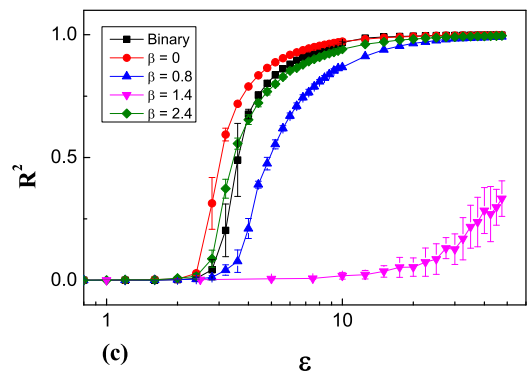}
\caption{(Color online) Under the WDU scheme, the variations of the
synchronization order parameter, $R^2$, as a function of the
coupling strength, $\varepsilon$, for scale-free networks (a) (c)
and small-world networks (b). As a reference, the synchronization of
non-weighted network (Binary) is also plotted in each subplot. All
the networks have size $N=6400$ and average degree $6$. In
constructing the WDU network, each link is arranged a weight chosen
randomly from the range $[1,100]$ for (a) and (b), and from the
range $[1,100000]$ for (c). The rewiring probability in generating
the small-world networks is $p_r=0.2$. The error bars are estimated
by the standard deviation, and each data is averaged over 20 network
runs.} \label{fig2}
\end{figure}

We start by demonstrating the typical path from non-synchronous to
synchronous states in non-weighted complex networks, i.e. the case
of $\alpha=0$ in Eq. (\ref{alpha}), so as to define the onset of
network synchronization. As plot in Fig. \ref{fig1} (the curve for
$\alpha=0$), when the coupling strength is small, e.g. $\varepsilon
\approx 0$, the trajectories of the oscillators are almost
uncorrelated, and we roughly have $R \approx O(1/\sqrt N)$ for the
order parameter. This value of $R$ keeps almost unchanged till a
critical coupling strength, $\varepsilon_c$, is met. After that, as
$\varepsilon$ increases from $\varepsilon_c$, the value of $R$ will
be gradually increased and, when $\varepsilon$ is large enough, it
reaches the limit $R=1$ (the state of complete network
synchronization). The value $\varepsilon_c$ is then defined as the
onset point for synchronization, which, for the case of non-weighted
scale-free network, is about $2$ (Fig. \ref{fig1}). In numerical
simulations, we define the onset of network synchronization as the
point where $R^2>2\times 10^{-2}$. In what follows, we will study
how the value of $\varepsilon_c$ is changed by the weight
parameters, $\alpha$ and $\beta$, in the two types of networks.

We first check the influence of the weight parameter, $\alpha$, on
WDC-type networks. The results are carried for scale-free networks
and are plotted in Fig. \ref{fig1}. It is shown that, as the weight
parameter increases from $-3$ to $5$, the network synchronization is
\emph{monotonically} enhanced. In particular, the point of onset of
synchronization, i.e. the vale of $\varepsilon_c$, is found to be
gradually shifted to the small values. For instance, when
$\alpha=-3$, the critical coupling strength is about $7$, while this
value is decreased to about $0.9$ when $\alpha=3$. Another
observation in Fig. \ref{fig1} is that, in the regime of
$\alpha<-1$, a small change of the weight parameter could induce a
significant change to the network synchronization; while in the
regime of $\alpha>-1$ the similar change in $\alpha$ has relatively
small influence. For instance, by increasing $\alpha$ from $-3$ to
$-2$, the critical coupling strength is decreased from $7$ to $4.1$;
while it is almost unchanged when $\alpha$ is increased from $3$ to
$5$ (see Fig. \ref{fig1}).

We next investigate the influence of the weight parameter on
WDU-type networks. In Fig. \ref{fig2}(a) and (c), we plot the
variation of the order parameter as a function of the coupling
strength for the scale-free network. It is found that, similar to
the case of WDC networks, the network synchronization is also
influenced by the weight parameter. The new finding is that, in
WDU-type networks the influence of the weight parameter on network
synchronization is \emph{non-monotonic}. In particular, it can been
seen from Fig. \ref{fig2}(a) and (c), as $\beta$ increases from
$0.8$ to $1.4$ the network synchronization is deteriorated; while
increasing $\beta$ from $1.4$ to $2.4$, the network synchronization
is enhanced. Speaking alternatively, as the weight parameter varies,
there is a \emph{crossover} in the synchronization performance for
WDU network. Under the same weighting scheme (WDU network), the
crossover of network synchronization is also observed in the model
of small-world networks, which can be seen from Fig. \ref{fig2}(b).
In particular, the max weight $c_{max}$ has no effect on the onset
of synchronization on scale-free networks and small world networks
(only give the examples on scale-free networks), qualitatively.

The results of Figs. \ref{fig1} and \ref{fig2} provide us with the
following information: 1) the weight parameters have important
effects on the network synchronization, and 2) the effects of the
weight parameters on network synchronization are dependent on the
weight-degree correlation. To explore these findings in depth, we go
on to study the synchronization mechanisms of the networks.

\section{The influence of weight-degree correlation on network synchronization}\label{criticalcouple}

In studying the dynamics of directed and weighted complex networks,
an effective method is to analyze the properties of the weight
gradient \cite{XGW:2007,XGW:2008}. For a directed network, the pair
of weights on a link in general is different from each other, which
causes a gradient (bias) among the connected nodes. The idea behind
gradient analysis is to extract the asymmetric part (gradient) on
each link, and analyze its functions to the network dynamics
separately. In this way, the original network can be virtually
regarded as a superposition of two subnetworks: one constituted by
the symmetric links and the other one constituted by gradient links.
In Ref. \cite{XGW:2007}, the later is also called the gradient
complex network.

Gradient network has been employed in analyzing the network
synchronization in literature. In Ref. \cite{XGW:2007} it is shown
that, with the increase of the gradient strength, the complete
synchronization in scale-free network can be significantly enhanced.
The influence of gradient on the onset of synchronization of coupled
phase oscillators has been also investigated \cite{XGW:Chaos}, where
an important finding is that, besides the gradient strength, the
gradient direction also plays a key role in influencing the network
synchronization. In particular, in Ref. \cite{XGW:Chaos} it is
demonstrated that when the gradient direction is pointing from the
larger-degree to smaller-degree nodes on each link, the increase of
the gradient strength will enhance the synchronization
monotonically; while if it is in the opposite direction, the
increase of the gradient strength will suppress the synchronization.
Noticing of these effects of the weight gradient and gradient
direction on the network synchronization, we in the following
explore the properties of the gradients in our models of WDC and WDU
networks, with a hope to explain the numerical observations in Figs.
\ref{fig1} and \ref{fig2}.

Following the method of Ref. \cite{XGW:2007}, we construct the
gradient networks in our models, as follows. The Eq.
(\ref{Kuramoto}) can be rewritten as
\begin{equation}
\dot{\phi}_m = \omega_m + \varepsilon \sum_{n=1}^N
g_{mn}\sin(\phi_n-\phi_m),\label{Kuramoto1}
\end{equation}
with $g_{mn}=c_{mn}/s_m$ the normalized coupling strength, which
captures the real coupling strength that node $m$ receives from node
$n$. The reason for the normalized coupling strength is simply for
convenience, since in this way the total network coupling strength
will be keeping as constant and is not changing with the weight
parameters. That is to say, in our models the change of network
synchronization is caused solely by the weight distribution, instead
of the change of the coupling cost.

The coupling gradient, $\Delta g_{mn}$, flowing from node $n$ to
node $m$ under the condition $k_n>k_m$ is written as
\begin{equation}
\Delta g_{mn}=
g_{mn}-g_{nm}=\frac{c_{mn}}{s_m}-\frac{c_{nm}}{s_n}\label{Deltag_{mn}},
\end{equation}
where $c_{mn}=c_{nm}$ is the weight before the normalization
operation, and $s$ is the node weight intensity defined in Sec. II.
Extracting only the gradient element on each of the network links,
together with the network nodes, we can construct the gradient
network. In our study, the positive direction of the gradient is
defined as pointing from the larger-degree to smaller-degree nodes
on each link. With this definition, we can measure the gradient at
the network level, i.e. the averaged network gradient, which is
  \begin{equation}
  \langle \Delta g\rangle =\frac{ \sum_{mn,k_m<k_n}\Delta g_{mn}}{M},
\end{equation}
where $\langle \ldots \rangle$ means that the value is averaged over
the network links (totally there are $2M=N\langle k\rangle$ links in
the network). Please note that the summation in Eq. (7) is under the
condition of $k_m<k_n$, which is based on the direction of coupling
gradient defined above. Our next job is to analyze how the tuning of
the weight parameter will change the gradient network and, in turn,
affect the network synchronization.

\subsection{WDC-type Networks}

We first analyze the synchronization of WDC network. For this type
of weighting scheme, the node weight intensity is
$s_m=k_m^\alpha\sum_{n\in \wedge_m} k_n^\alpha =
k_m^{1+\alpha}\langle k_n^\alpha\rangle_m$. According to our
definitions above, the coupling gradient between the pair of
connected nodes, $m$ and $n$, is,
\begin{equation}
\Delta g_{mn}=\frac{c_{mn}}{s_m}-\frac{c_{nm}}{s_n}=\frac{k_m^\alpha
k_n^\alpha}{k_m^{1+\alpha}\langle
k_l^\alpha\rangle_m}-\frac{k_m^\alpha
k_n^\alpha}{k_n^{1+\alpha}\langle
k_l^\alpha\rangle_n}.\label{gradient}
\end{equation}
For scale-free networks generated by the BA standard model (the
model we have adopted for WDC network in Fig. \ref{fig1}), the
degrees of the nodes are not correlated, and we can roughly treat
$\langle k_n^\alpha\rangle _m$ as a constant, $\langle
k_n^\alpha\rangle _m=(1/k_m)\sum k_n^\alpha$ \cite{JYA:2007}. Using
this approximation, the above equation can be simplified into
\begin{equation}
\Delta g_{mn}\sim \frac{k_m^\alpha
k_n^\alpha}{k_m^{1+\alpha}}-\frac{k_m^\alpha
k_n^\alpha}{k_n^{1+\alpha}} =
\frac{k_n^{1+\alpha}}{k_mk_n}-\frac{k_m^{1+\alpha}}{k_mk_n}.\label{gradient1}
\end{equation}
As shown by the above equation, the coupling gradient arises
naturally when one of the conditions, $\alpha\neq -1$ or $k_m\neq
k_n$, is satisfied, which is generally met in weighted complex
network.

Separated by the value $\alpha=-1$, the function of the gradient at
the two sides of the parameter space is completely different.
Assuming that $k_m<k_n$ in Eq. (\ref{gradient1}), in the region of
$\alpha<-1$, we statistically have $s_m>s_n$ and, consequently,
$\Delta g_{mn}<0$. Since $s_m \propto k_m$, the gradient thus is
pointing from the smaller-degree to larger-degree nodes. According
to the result of Ref. \cite{XGW:Chaos}, this negative gradient will
suppress synchronization and, moreover,  in the negative regime with
the increase of the gradient strength, $\langle \Delta g_{mn}
\rangle$, the value of $\varepsilon_c$ will be monotonically
decreased. In a similar fashion, in the region of $\alpha>-1$, the
gradient is pointing from the larger-degree to smaller-degree nodes,
i.e. the gradient is positive. In this case, with the increase of
the gradient strength, the value of $\varepsilon_c$ will be also
monotonically decreased. Combining the behaviors of $\varepsilon_c$
in the two regimes, we thus predict a continuous decrease of the
critical coupling strength as a function of the weight parameter,
$\alpha$, in the WDC network.

\begin{figure}[tbp]
\includegraphics[width=0.85\linewidth]{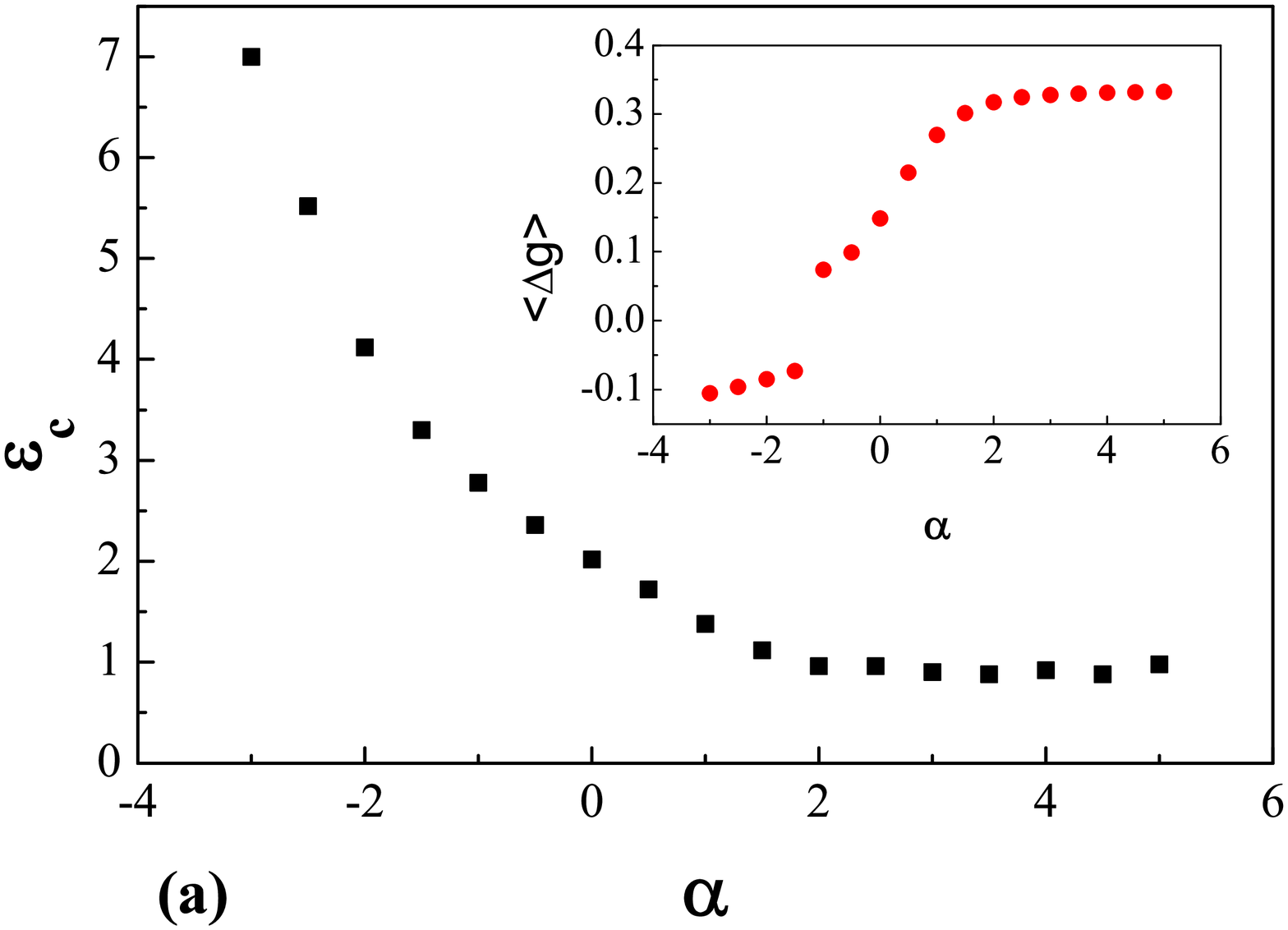}
\includegraphics[width=0.85\linewidth]{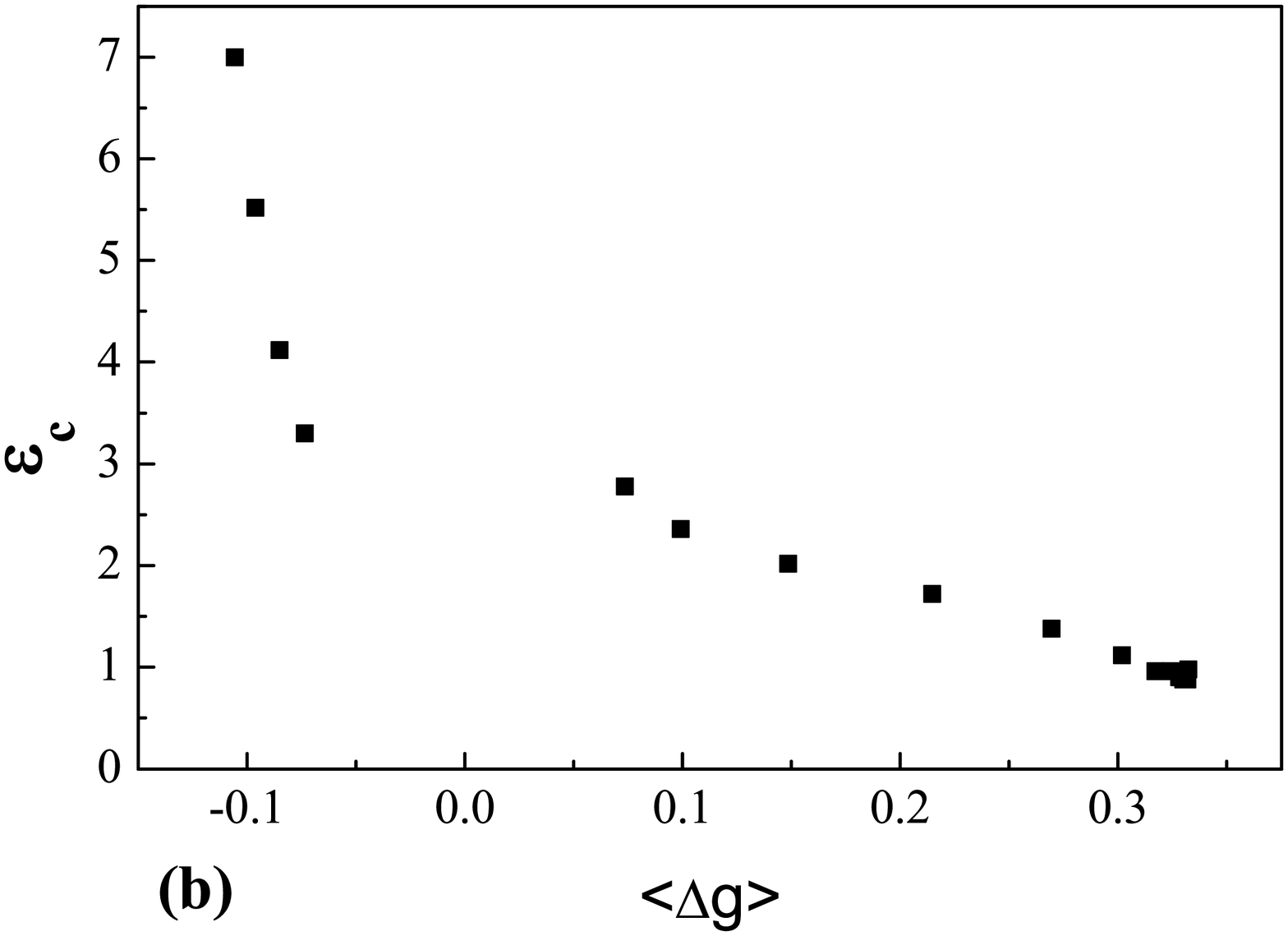}
\caption{(Color online) For the same scale-free network as used in
Fig. 1, under WDC weighting scheme, the variation of the critical
coupling strength, $\varepsilon_c$, as a function of the weight
parameter, $\alpha$, in (a) and the average network gradient,
$\langle \Delta g\rangle$, in (b). Inset: $\langle \Delta g\rangle$
versus $\alpha$. Each data is averaged over 20 network runs.}
\label{fig3}
\end{figure}

The above predictions are verified by numerical simulations. By the
scale-free network used in Fig. \ref{fig1}, we plot the variation of
$\varepsilon_c$ as a function of $\alpha$ in Fig. \ref{fig3}(a).
Clearly, with the increase of $\alpha$ the network synchronization
is found to be monotonically enhanced (a continuous decrease of
$\varepsilon_c$). In the inset of Fig. \ref{fig3}(a), we also plot
the variation of the averaged network gradient, $\langle \Delta g
\rangle$, as a function of $\alpha$, where a continuous increase of
$\langle \Delta g \rangle$ is evident. The enhanced synchronization
by coupling gradient is more clearly shown in Fig. \ref{fig3}(b), in
which the value of $\varepsilon_c$ is found to be monotonically
decreased with the increase of the averaged network gradient.

Our findings in WDC network, i.e., the results in Fig. \ref{fig3},
might give indications to the design of realistic networks where
synchronization is importantly concerned, e.g., the world-wide
airport network \cite{PNAS_101_3747} and the metabolic network
\cite{EPL_72_308}. In a practical situation, due to the high cost of
maintaining the weight gradient, a tradeoff between the network
performance (synchronization) and the weight (coupling) gradient
will be necessary. The results of Fig. \ref{fig3}(a) suggest that in
the region of $\alpha<2$, the network synchronization can be
\emph{significantly} improved by a small gradient cost (small change
of $\alpha$); while in the region of $\alpha>2$, to achieve the same
performance improvement, the gradient cost could be significantly
large. This might be one of the reasons why some realistic networks
are found to be possessing weight parameter $\alpha\approx 0.5$,
e.g., the world-wide airport network \cite{PNAS_101_3747} and the E.
coli metabolic network \cite{EPL_72_308}.

\subsection{WDU-type Networks}

\begin{figure}
\includegraphics[width=0.85\linewidth]{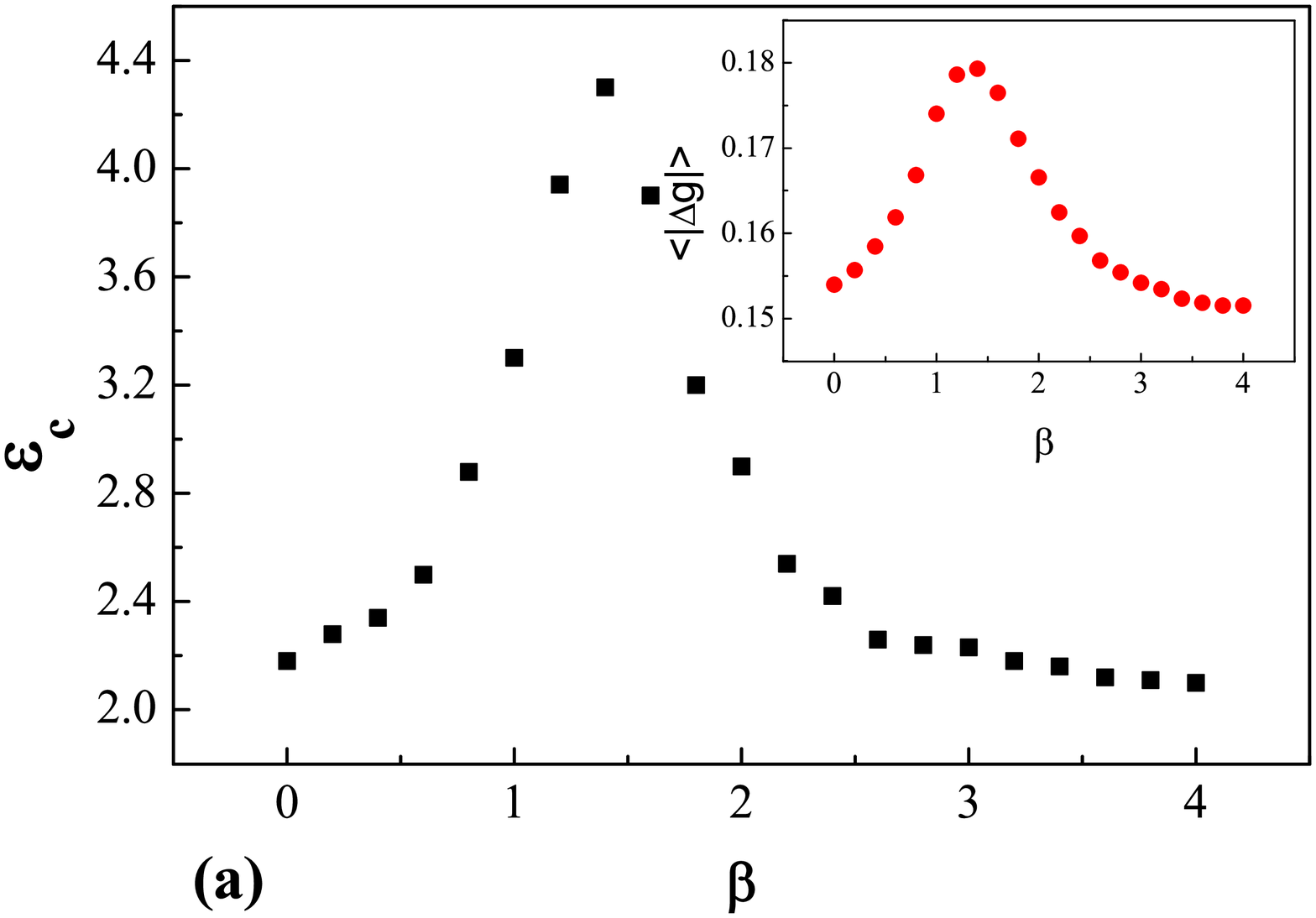}
\includegraphics[width=0.85\linewidth]{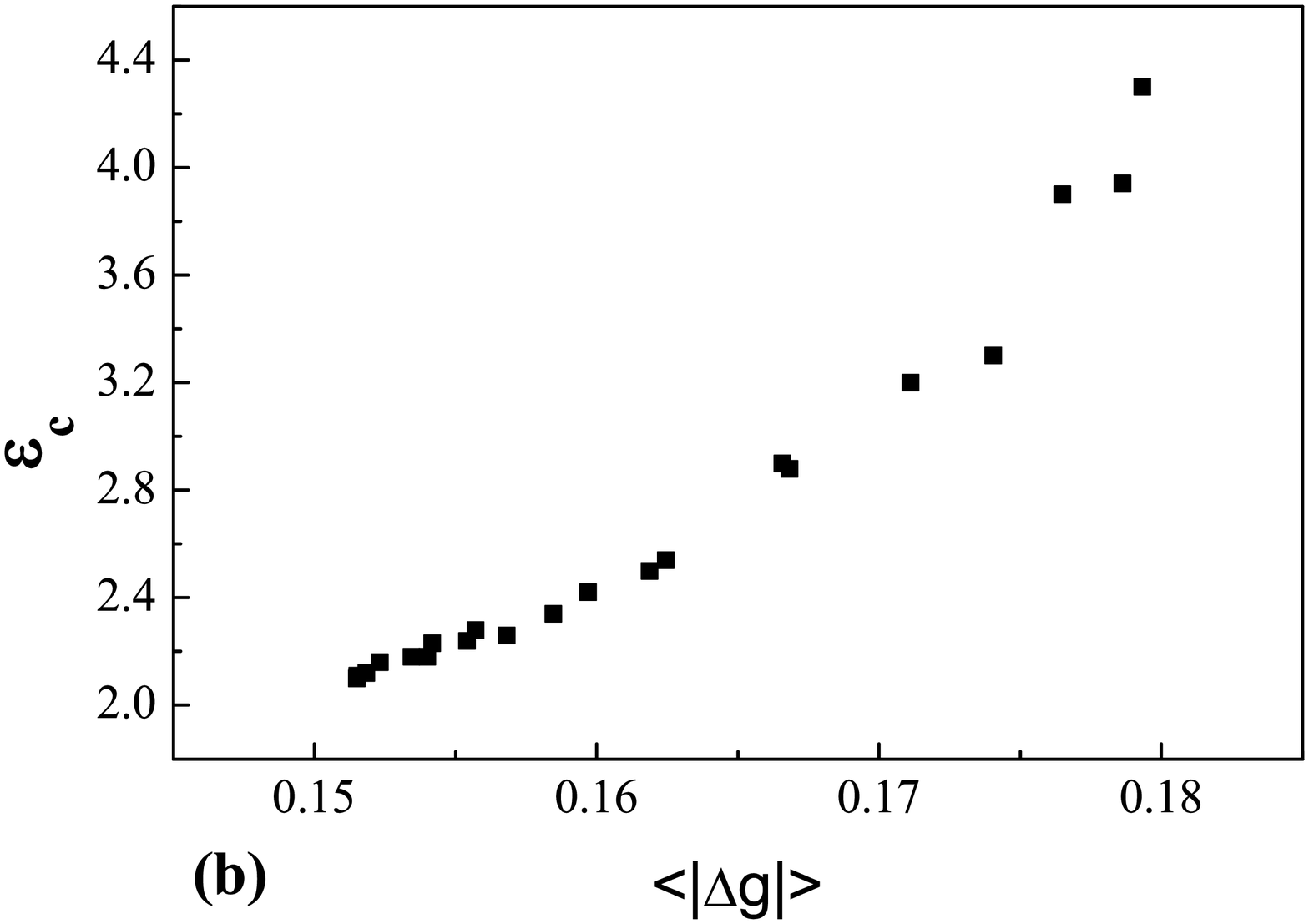}
\caption{(Color online) For the same scale-free network as used in
Fig. 2(a), under WDU scheme, the variation of the critical coupling
strength, $\varepsilon_c$, as a function of the weight parameter,
$\beta$, in (a) and the averaged network gradient, $\langle |\Delta
g|\rangle$, in (b). Insets: $\langle |\Delta g|\rangle$ versus
$\beta$. Each data is averaged over 20 network runs.} \label{fig4}
\end{figure}

We next analyze the synchronization of WDU network. As indicated in
Fig. \ref{fig2}, in this type of networks with the increase of the
weight parameter, $\beta$, the network synchronization may be either
enhanced or suppressed. To have a close look to the relationship
between $\varepsilon_c$ on $\beta$, we increase $\beta$ from $0$ to
$4$ gradually, and checking the variation of $\varepsilon_c$. The
numerical results carried on scale-free networks are plotted in Fig.
\ref{fig4}(a). In this figure, it is clearly shown that at about
$\beta_c\approx 1.5$, there exists a maximum value for
$\varepsilon_c$. It is worthwhile to note that the cap-shape
variation of $\varepsilon_c$ may have implication to the function of
some realistic networks where network desynchronization
\cite{DESYN}, instead of synchronization, is desired. Say, for
instance, in the mobile communication network, simultaneous
communications may cause the information congestion and should be
always avoided. An interesting empirical observation is that, in
this weighted network, the weight of the network links follows
roughly a pow-law distribution with an exponent $\beta\approx 1.9$
\cite{PNAS_104_7332}. This value is quite close to $\beta_c$ in Fig.
\ref{fig4}(a), which might indicate the possible application of
weight-degree correlation in affecting network dynamics for this
specific network.

We now analyze the synchronization behavior of WDU network from the
viewpoint of gradient network. It should be noted that, different to
the case of WDC network, in WDU network the direction of the
gradient is not uniquely defined. That is, the gradients on the
network links may either pointing from the larger-degree node to the
smaller-degree node, or pointing in the opposite direction. As a
matter of fact, in WDU network the gradient directions are
determined by the weight details, instead of node degrees. That is,
the positive and negative gradients are coexisting in a WDU network.
This arises a problem in using the averaged network gradient for the
synchronization analysis, since for a large-scale WDU network, due
to the random weight arrangement, we roughly have $\langle \Delta g
\rangle \approx 0$. Regarding of this, in WDU network we use the
averaged gradient amplitude, $\langle |\Delta g| \rangle =\frac{
\sum_{mn}|\Delta g_{mn}|}{M}$, to characterize the gradient at the
network level. With this adoption, our task now is shifted to
exploring the relationship between $\varepsilon_c$ and $\langle
|\Delta g| \rangle$.

\begin{figure}
\includegraphics[width=0.85\linewidth]{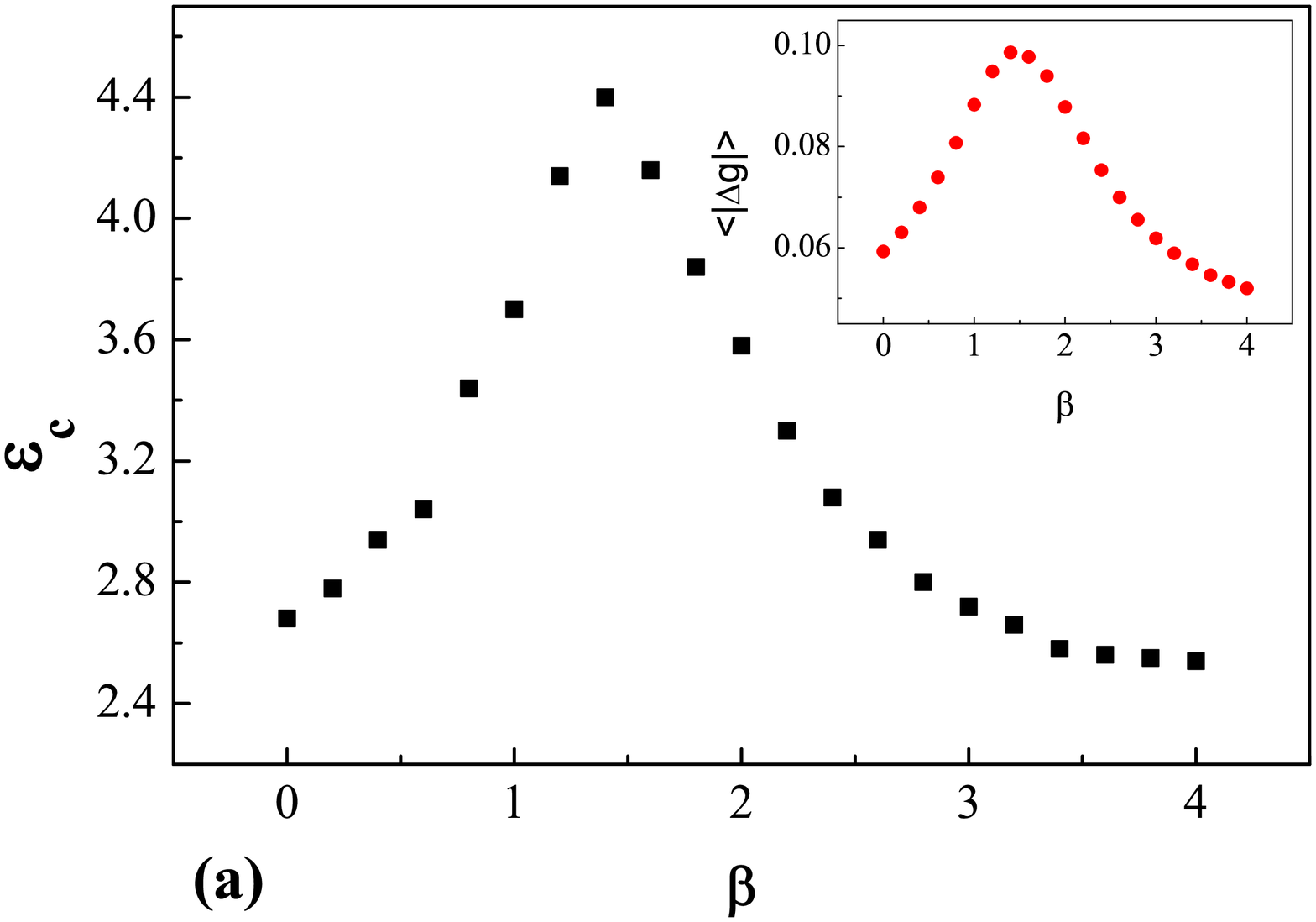}
\includegraphics[width=0.85\linewidth]{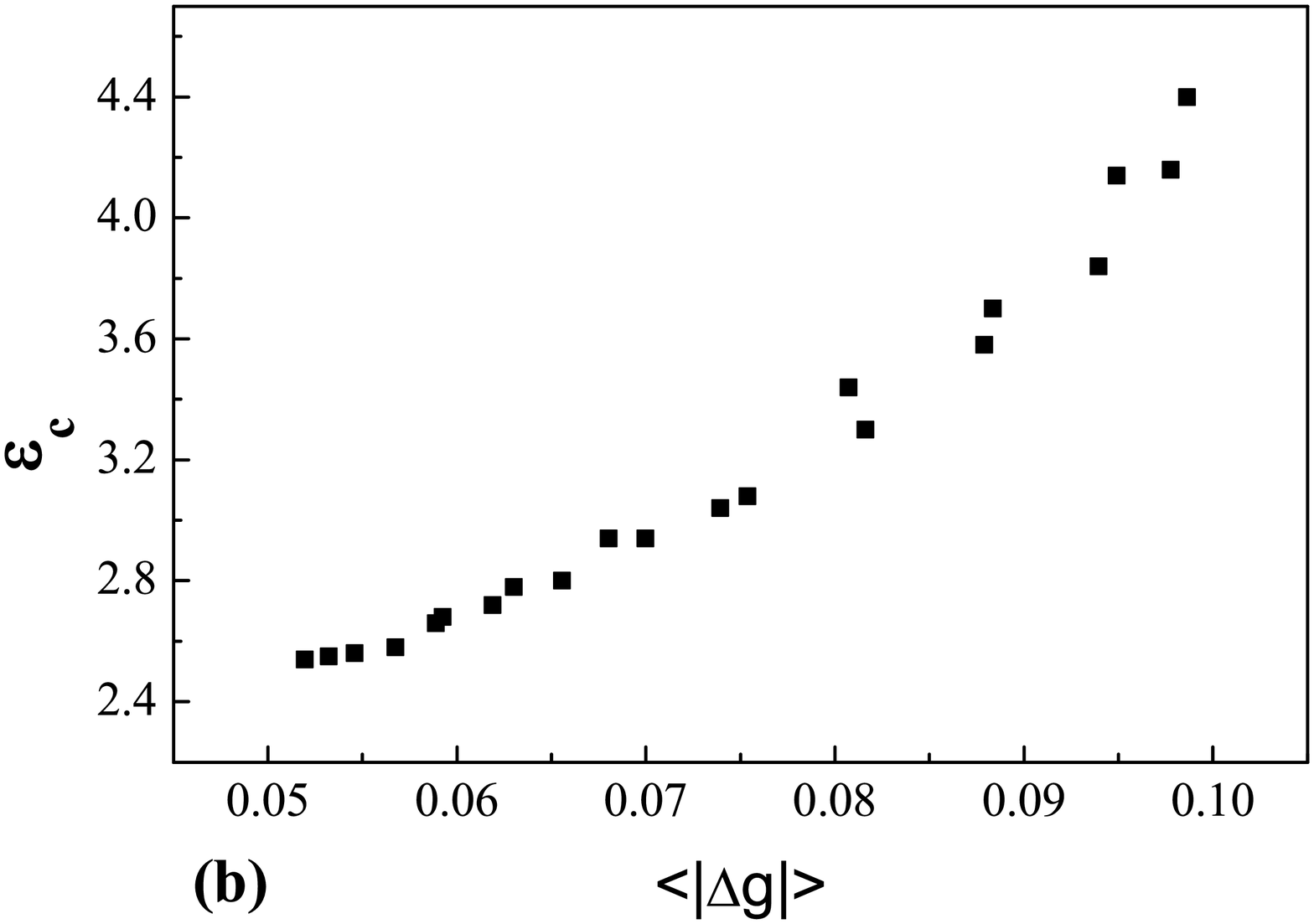}
\caption{(Color online) For the same small-world network as used in
Fig. 2(b), under WDU scheme, the variation of the critical coupling
strength, $\varepsilon_c$, as a function of the weight parameter,
$\beta$, in (a) and the average network gradient, $\langle |\Delta
g|\rangle$, in (b). Insets: $\langle |\Delta g|\rangle$ versus
$\beta$. Each data is averaged over 20 network runs.} \label{fig5}
\end{figure}

We first check the variation of $\langle |\Delta g| \rangle$ as a
function of $\beta$. The results are plotted in the inset of Fig.
\ref{fig4}(a). Different to the case of WDC network, it is observed
that here the variation of $\langle |\Delta g| \rangle$ has a
cap-shape structure. (For WDC network, as $\alpha$ increases from
$0$, the value of $\langle \Delta g \rangle$ is monotonically
increased, see Fig. \ref{fig3}(a).) In particular, at the critical
parameter $\beta_c$, we also find a maximum value for $\langle
|\Delta g| \rangle$. The similar behavior of $\varepsilon_c$ and
$\langle |\Delta g| \rangle$ seems to indicate the following:
\emph{gradient deteriorates synchronization in WDU network}. To
check this out, we plot in Fig. \ref{fig4}(b) the dependence of
$\varepsilon_c$ on $\langle |\Delta g| \rangle$. Now it is clearly
seen that, as the network gradient increases, the network
synchronization is monotonically suppressed.

The suppressed synchronization by gradient is also found in
small-world networks under the WDU scheme. With the small-world
network model used in Fig. \ref{fig2}, we calculate the variation of
the critical coupling strength as a function of the weight
parameter. The results are presented in Fig. \ref{fig5}(a). Again,
the maximum value of $\varepsilon_c$ is found at some intermediate
value of $\beta$. Similar to the case of scale-free network (Fig.
\ref{fig4}), the averaged gradient amplitude, $\langle |\Delta g|
\rangle$, also shows a cap-shape structure (inset of Fig.
\ref{fig5}(a)), with the maximum value of $\langle |\Delta g|
\rangle$ appears at $\beta_c$. The relationship between
$\varepsilon_c$ and $\langle |\Delta g| \rangle$ is plotted in Fig.
\ref{fig5}(b), where the suppression of synchronization by gradient
is evident.

\subsection{Mechanism analysis}

Why synchronization is suppressed by gradient in WDU network, while
is enhanced in WDC network? To answer this, we compare the
structures of the gradient networks generated in the two types of
networks, and analyzed their functions to the network
synchronization. For a WDC network, according to the study of Ref.
\cite{XGW:2008}, the gradient couplings of the network are organized
into a spanned tree. More specifically, at the top of the gradient
network there locates the largest-degree node of the network, which
sending out gradients to its nearest neighbors (the second level).
Each node at the second level is then sending out gradients to its
own nearest neighbors (the third level), given that it has a larger
degree to its neighbors. In this way, all the network nodes will be
organized into a one-way coupled tree structure. An important
feature of this gradient tree is that, starting from the rooting
node (the largest-degree node), every node of the network can be
reached by following the gradient couplings. Previous studies on
gradient network have shown that \cite{XGW:2007}, for such a spanned
gradient tree, the network synchronizability will be monotonically
increased with the increase of the gradient strength. From Eqs.
(\ref{alpha}) and (\ref{Deltag_{mn}}), it is straightforward to see
that in our model of WDC network, the gradient strength is
increasing with the parameter $\alpha$ in a monotonic fashion. We
thus attribute the enhanced synchronization in WDC network to the
existence of a spanned gradient tree.

\begin{figure}
\includegraphics[width=0.9\linewidth]{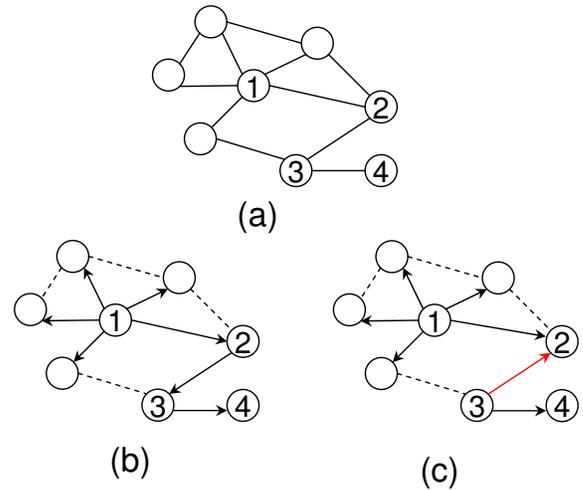}
\caption{(Color online) A schematic plot on the distribution of the
gradient couplings in WDC and WDU networks. (a) The simplified model
of undirected, non-weighted network. (b) The distribution of the
gradient couplings under the WDC-scheme. The network nodes and the
gradient couplings are organized into a spanned gradient tree, in
which each node is reachable from the rooting node (numbered 1). (c)
The distribution of the gradient couplings under the WDU-scheme, in
which the weight on the link between nodes $3$ and $4$ is increased
by $\Delta$, while the remaining gradients keep identical to that of
(b). As $\Delta$ exceeds some critical value, the gradient between
nodes 2 and 3 will switch its direction (the red arrow-line),
manifesting a breaking of the gradient network. Dashed lines are
links dominated by symmetric couplings (with negligible gradient).}
\label{fig6}
\end{figure}

To better describe the spanned gradient tree in WDC network, we
adopt the simplified network model plotted in Fig. \ref{fig6}(a),
which might be also regarded as a motif of a large-scale network.
According to our model (the Eq. (\ref{Kuramoto})), under WDC scheme
the gradient direction will be pointing from larger-degree node to
smaller-degree node on each link. Following this rule, we can
construct the gradient network, which is schematically shown in Fig.
\ref{fig6}(b). Please note that, to illustrate, here we only show
the gradient direction, while neglecting the gradient strength. In
Fig. \ref{fig6}(b), it is clearly seen that, leading by node $1$,
all the network nodes are connected by the gradient couplings in a
hierarchical fashion. It is important to note that, while the exact
gradient strength of each link is modified by the parameter
$\alpha$, the hierarchical structure of the spanned gradient tree is
unchanged. In particular, in the gradient network node $4$ is always
reachable by node $1$, by passing the nodes $2$ and $3$.

We go on to study the gradient couplings in WDU network. To manifest
the feature of weight-degree non-correlation, and also for the sake
of simplicity, we keep all other weights of the network links the
same to those of Fig. \ref{fig6}(b) (WDC network), while changing
only the weight between nodes $3$ and $4$ by a large increment,
$\Delta$. Taking into account this single change of the link weight,
we now reanalyze the gradient couplings of the network. Since all
other weights of the network are unchanged, all the gradients are
the same to that of Fig. \ref{fig6}(b), except the one on the link
between nodes $2$ and $3$. According to our model of Eq.
(\ref{Kuramoto}), the gradient from node $2$ to node $3$ is $\Delta
g_{23} = c_{23}(1/s_3-1/s_2)$, with $s$ the weight intensity defined
in Sec. II. In WDC network (Fig. \ref{fig6}(b)), because $s_2>s_3$,
we thus have $\Delta g_{23}>0$. This direction of gradient, however,
will be switched in WDU network if $\Delta$ is large enough (Fig.
\ref{fig6}(c)). In the WDU network, the gradient from node $2$ to
node $3$ is $\delta g'_{23} = c_{23}(1/s'_3-1/s_2)$, with
$s'_3=s_3+\Delta$. It is not difficult to find that, as long as
$\Delta > s_2-s_3$, the gradient will be pointing from node $3$ to
node $2$. That is, the gradient direction is \emph{switched}. Once
switched, the gradient network will be separated into two sub-trees:
one led by node $1$ and the other one led by node $3$. Previous
studies have shown that \cite{XGW:2008}, for such a breaking
network, the network synchronizability will be monotonically
deteriorated by increasing the gradient strength. Based on this
analysis, we attribute the deteriorated synchronization in WDU
networks to the breaking of the gradient network.

\section{Discussion and Conclusion}\label{conclud}

The main purpose of the present paper is to highlight the important
role of the weight-degree correlation played on the network
synchronization. While synchronization in weighted complex network
has been extensively studied in literature, to the best of our
knowledge, it is the first time that the correlation between the
link weight and node degree be investigated. To demonstrate the
effect of weight-degree correlation on the network dynamics, we have
employed the generalized Kuramoto model as the platform and,
following the tradition, adopting the scheme of normalized coupling
strength. The main findings of the paper, however, are general and
are expected to be observable for other network models and coupling
schemes as well, say, for instance, the complete network
synchronization.

Motivated by the empirical observations, we have adopted two
different weighting schemes for the WDC and WDU networks. The weight
parameters, $\alpha$ and $\beta$, thus have different meanings and,
in general, should not be directly compared. For this regard, the
comparison we have made in the present work is actually on the
\emph{different dynamical responses} of the two types of networks to
the change of the weight heterogeneity, which should not be
understood as a comparison of the network synchronizability. Direct
comparison of the two weighting schemes might be possible for the
BA-type scale-free networks, in which we roughly have
$\beta=-\gamma/(2\alpha)$. However, to keep this relation, in
numerical simulations the network size should be very large (larger
than $1\times 10^6$), which is out of our current computational
ability.

While gradient coupling has been identified as important for network
synchronization, previous studies have been mainly concentrated on
the type of WDC network \cite{WXWang:2009}. As such, a general
observation is that, by increasing the gradient strength, the
network synchronizability is monotonically increased. This
observation, as we show in the present paper, is invalid for the
case of WDU network. In particular, in WDU network we find that,
instead of enhancement, the increase of the gradient could actually
\emph{suppress} synchronization under some conditions. This finding
could be helpful to the design of realistic networks which,
depending on the network functionality, may either in favor of
network synchronization or network desynchronization. Now, besides
the aspect of link weight, to manipulate the network
synchronization, we can also adjust the correlation between the link
weight and the node degree.

We wish to note that, despite of the recent progresses in the onset
of synchronization of networked phase oscillators, at the present
stage we are still short of a theory to predict precisely the point
of onset of synchronization \cite{ROH:2005}. This is specially the
case when the link weight is of heterogeneous distribution, e.g. the
WDC and WDU networks discussed here. While our main findings are
based on numerical simulations, it is our believing that these
findings are general and stand for other network models and dynamics
as well.

In summary, we have investigated the onset of synchronization in
weighted complex networks, and explored the influence of
weight-degree correlation on the network synchronization. An
interesting finding is that, as the weight parameter increases, in
WDC network the synchronization is monotonically enhanced, while in
WDU network the synchronization could be either enhanced or
deteriorated. This finding is obtained from extensive simulations,
and is explained from the viewpoint of gradient network. Our study
shows that, besides the link weight, the weight-degree correlation
is also an important concern of the network dynamics.

X. Wang is supported by NSFC under Grant No. 10805038 and by Chinese
Universities Scientific Fund. Y. Fan and Z. Di are supported by 985
Project and NSFC under Grant No. $70771011$ and No. $60974084$. This
work is also supported by DSTA of Singapore under Project Agreement
POD0613356.

\end{document}